\documentclass[11pt,a4paper]{article}

\usepackage{jcappub}
\usepackage{graphicx}
\usepackage{amsmath,amsfonts,amssymb}
\usepackage{hyperref}
\usepackage{color}
\usepackage{verbatim}

\def\lsim{\mathrel{\raise.3ex\hbox{$<$\kern-.75em\lower1ex\hbox{$\sim$}}}}
\def\gsim{\mathrel{\raise.3ex\hbox{$>$\kern-.75em\lower1ex\hbox{$\sim$}}}}

\newcommand{\be}{\begin{equation}}
\newcommand{\ee}{\end{equation}}
\newcommand{\bea}{\begin{equation}\begin{aligned}}
\newcommand{\eea}{\end{aligned}\end{equation}}
\newcommand{\td}{{\rm d}}
\newcommand{\Msun}{M_{\odot}}

\newcommand{\Gpc}{{\rm Gpc}}
\newcommand{\yr}{{\rm yr}}

\newcommand{\eg}{\emph{e.g.}}
\newcommand{\ie}{\emph{i.e.}}

\title{Two populations of LIGO-Virgo \\ black holes}

\author[a]{Gert H\"utsi,}
\author[a]{Martti Raidal,}
\author[b]{Ville Vaskonen,}
\author[a]{and Hardi Veerm\"{a}e}
\affiliation[a]{Laboratory for High Energy and Computational Physics, NICPB, R\"{a}vala 10, 10143 Tallinn, \\ Estonia}
\affiliation[b]{Institut de Física d'Altes Energies (IFAE), The Barcelona Institute of Science and Technology, Campus UAB, 08193 Bellaterra (Barcelona), Spain}  
\emailAdd{gert.hutsi@to.ee}
\emailAdd{martti.raidal@cern.ch}
\emailAdd{vvaskonen@ifae.es}
\emailAdd{hardi.veermae@cern.ch}

\abstract{We analyse the LIGO-Virgo data, including the recently released GWTC-2 dataset, to test a hypothesis that the data contains more than one population of black holes. We perform a maximum likelihood analysis including a population of astrophysical black holes with a truncated power-law mass function whose merger rate follows from star formation rate, and a population of primordial black holes for which we consider log-normal and critical collapse mass functions. We find that primordial black holes alone are strongly disfavoured by the data, while the best fit is obtained for the template combining astrophysical and primordial merger rates. Alternatively, the data may hint towards two different astrophysical black hole populations. We also update the constraints on primordial black hole abundance from LIGO-Virgo observations finding that in the $2-400\Msun$ mass range they must comprise less than 0.2\% of dark matter.}
\begin{document}

\maketitle

\section{Introduction}
\label{sec:intro}

The first direct measurement of gravitational waves (GWs) by the LIGO-Virgo Collaboration~\cite{Abbott:2016blz} established gravitational wave astronomy. So far, the results have already provided additional confirmation that with good accuracy all gravitational observables can be described by General Relativity~\cite{LIGOScientific:2019fpa,Abbott:2020jks}. The recently released GWTC-2 dataset~\cite{Abbott:2020niy}, including the data from the first half of the third observing run (O3a), more than quadruples the collected LIGO-Virgo black hole (BH) merger events compared to the first two runs~\cite{LIGOScientific:2018mvr}, allowing for statistical analyses of BH binary physics~\cite{Abbott:2020gyp}. In the light of improved statistics, it is natural to ask whether the new data could shed light on the presently unknown origin of the observed BH binaries.

As the very first observed mergers involved BHs which were more massive than astrophysicists anticipated at that time -- the known masses of BHs in Galactic X-ray binaries are in a range $\sim 5-15\,\Msun$~\cite{Remillard:2006fc} -- an immediate reaction was that the progenitors of the LIGO-Virgo GW events might have a primordial origin~\cite{Sasaki:2016jop,Bird:2016dcv,Clesse:2016vqa}. This proposal is particularly interesting because primordial black holes (PBHs) could constitute the dark matter (DM) of the Universe. However, the observed merger rate alone constrains the allowed ${\cal O}(30) \Msun$ PBH DM abundance to be marginal~\cite{Sasaki:2016jop,Raidal:2017mfl,Ali-Haimoud:2017rtz}\footnote{For a recent review on other constraints on PBH abundance see, \eg,~Ref.~\cite{Carr:2020gox}.}. Dedicated simulations of PBH evolution in the early Universe~\cite{Raidal:2018bbj,Inman:2019wvr}, studies of the formation of small-scale structure~\cite{Hutsi:2019hlw,Vaskonen:2019jpv,Gow:2019pok,DeLuca:2020jug}, as well as several other analyses~\cite{Liu:2018ess,Liu:2019rnx,DeLuca:2020qqa,Hall:2020daa,DeLuca:2020fpg,DeLuca:2020sae} have confirmed this result. Nevertheless, if the PBHs comprise just a fraction of DM, the primordial hypothesis for the LIGO-Virgo GW events is still physically interesting and motivates further studies of the PBH formation and evolution in the early Universe.

It was soon realized that the observed LIGO-Virgo BHs can easily be astrophysical since the stellar evolution studies show that the observed ${\cal O}(30) \Msun$ BHs can be formed in the collapse of metal-poor stars (see, \eg,~Ref.~\cite{2002RvMP...74.1015W}). In addition, it is likely that there exists a mass gap for heavier astrophysical black holes (ABHs)~\cite{1967ApJ...148..803R,1968Ap&SS...2...96F,2002RvMP...74.1015W}, and the LIGO-Virgo data seems to be consistent with this hypothesis~\cite{Fishbach:2017zga}. There are several astrophysical mechanisms for BH formation -- the two main channels being isolated (or `field') binary evolution and dynamical capture in dense stellar environments -- thus there could be several different astrophysical populations of merging BHs. Therefore the increase of LIGO-Virgo statistics is important for these studies as well. For example, the first investigations using the new GWTC-2 catalog seem to indicate that a mixture of `isolated' and `dynamical' channels is necessary to fit the data. However, there is currently significant disagreement about their relative contributions: Ref.~\cite{Zevin:2020gbd} finds that $\sim 90\%$ come from isolated binary evolution channel whereas Ref.~\cite{Wong:2020ise} claims that the majority, $\sim 80\%$, is provided by dynamical capture. 

The aim of this work is to search for possible indications for several populations of BHs in the LIGO-Virgo data. In this search, it is important to account for the possibility that different fractions of the events may have a different origin. In particular, we concentrate on the possibility that the astrophysical and primordial populations may coexist. 

The astrophysical binary BH population has several distinguishing features compared to the possible PBH population. Two features are particularly important for this study. First, independently of the specific formation mechanism of PBHs, they start to form binaries even before the matter-radiation equality. Therefore, whereas the merger rate of ABHs, which follows the star formation rate, drops at redshifts higher than $z \approx 2$, the PBH merger rate monotonically increases as a function of $z$ (up to $z\sim 1000$). However, as the LIGO-Virgo detectors are not sensitive to BH mergers at redshifts $z\gsim 1$, this difference between the astrophysical and primordial BH merger rates can not be probed yet~\cite{Fishbach:2018edt,Abbott:2020gyp}. Second, the PBH mass distribution is typically described by a smooth peak, whereas the ABH mass distribution can contain sharp cuts or gaps due to the specifics of ABHs formation. Since the GWTC-1 and GWTC-2 datasets contain 56 BH-BH merger events, we can start looking for features in the mass distribution of those BHs that can provide hints for different BH populations in the data. Already the GWTC-1 dataset showed a preference for scenarios containing only ABH over scenarios with PBH only~\cite{Hall:2020daa}. However, scenarios including both astrophysical and primordial BH binary populations have not been studied. As the LIGO-Virgo Collaboration analysis~\cite{Abbott:2020gyp} of the observed binary population prefers mass distribution templates with a peaked feature, albeit only at a $1.4\sigma$ CL, our study is consistent with, and partly motivated by, this result. 

The third distinguishing feature between astrophysical and primordial BHs is their spin distribution. Even though the LIGO-Virgo events taken separately have very poor sensitivity to BH spins, the cumulative data starts to provide some indicative information about the population-wide characteristics~\cite{Abbott:2020gyp,Wong:2020yig}. The initial spins of PBHs are expected to be tiny~\cite{Chiba:2017rvs,Mirbabayi:2019uph,DeLuca:2019buf} since they form at the highest density peaks in the primordial plasma that tend to have a spherical symmetry~\cite{Bardeen:1985tr}. At the same time, spins of the astrophysical binaries formed in the field have complex dependency on stellar winds and mass transfer and are mostly expected to be aligned with the orbital angular momentum, with possible misalignments caused by the momentum kicks imparted to the orbit during the core collapse~\cite{Kalogera:1999tq,Gerosa:2018wbw}. PBHs can build up their spins during subsequent evolution by accreting baryonic matter~\cite{Fernandez:2019kyb,DeLuca:2020qqa,DeLuca:2020bjf,DeLuca:2020fpg,Garcia-Bellido:2020pwq}. If PBHs are not the only form of DM, which they cannot be in the solar mass range~\cite{Carr:2020gox}, then PBHs are decorated by compact DM halos~\cite{Ricotti:2007au,Adamek:2019gns} which deepen the effective potential well and enhance the accretion rate. 

Accretion physics is very complex. Reliable estimates of the effects of accretion on PBH spins are thus difficult to obtain. Even the growth of an isolated BH has several large uncertainties, such as whether the geometry of the flow is spherical or disk-like, and how to reliably treat structure formation and baryonic feedback, \eg, X-ray heating, gas ionization, etc. The situation is even more uncertain for binary systems. A sufficiently compact binary can be contained inside its total systemic Bondi radius, and thus the accretion flow at far regions resembles that of a single accretor. The flow close to the binary is, however, much more complex. Fast-moving BHs scatter gas particles very efficiently and carve out a cavity around the binary~\cite{1994ApJ...421..651A}. Naively, this could significantly suppress accretion. However, in the case of disk accretion narrow accretion streamers will effectively carry gas from the inner parts of the circumbinary disk to the mini-disks surrounding BHs (see, \eg,  Refs.~\cite{Farris:2013uqa,Ragusa:2016qhp,Munoz:2018tnj}). Depending on the effective disk viscosity, the total efficiency of gas transport can vary significantly (see, \eg, Ref.~\cite{Ragusa:2016qhp,Munoz:2018tnj}). Considering the large uncertainties related to accretion physics, and expectations that at relatively low BH masses, ${\cal O}(10)\Msun$, accretion is not expected to grow the mass significantly, we do not include the possible PBH spins and their mass evolution in our analysis. Similarly, an improved understanding of the evolution of astrophysical BH binaries is needed to derive reliable quantitative results from the BH spin information. Therefore our analysis deals only with the mass and redshift distributions of the LIGO-Virgo BH binaries.


This paper is structured as follows. In Sec.~\ref{sec:mergers} we study the merger rate of PBH binaries. In Sec.~\ref{sec:implications} we perform a maximal likelihood analysis of the PBH models with log-normal and critical collapse mass functions and derive constraints on the PBH abundance. In Sec.~\ref{sec:discussion} we discuss astrophysical BH binary scenarios and, assuming a simplified model for the ABH merger rate, repeat the maximal likelihood analysis in scenarios containing both primordial and astrophysical BH binaries. We summarize our conclusions in Sec.~\ref{sec:concl}.

\section{PBH merger rate}
\label{sec:mergers}

The PBH binary formation mechanisms can be classified by the cosmological epoch in which the binaries form. First, slightly before the matter-radiation equality close PBH pairs decouple from the Hubble flow and form binaries~\cite{Nakamura:1997sm,Ioka:1998nz,Nakama:2016gzw,Sasaki:2016jop,Raidal:2017mfl,Ali-Haimoud:2017rtz,Raidal:2018bbj,Kavanagh:2018ggo,Vaskonen:2019jpv}. Second, at late times well after matter-radiation equality PBHs can form binaries from close encounters~\cite{1989ApJ...343..725Q,Mouri:2002mc,Bird:2016dcv,Clesse:2016vqa}. The PBH binaries formed in the early Universe dominate the present merger rate~\cite{Raidal:2017mfl,Ali-Haimoud:2017rtz,Sasaki:2018dmp,DeLuca:2020jug}, even though most of these binaries are disrupted by interactions with other PBHs before they merge~\cite{Raidal:2018bbj,Vaskonen:2019jpv}.
We will therefore focus on the merger rate of PBH binaries formed in the early Universe.

Before the PBHs decouple from expansion in the early Universe, they are distributed uniformly in space~\cite{Ali-Haimoud:2018dau}, and their peculiar velocities vanish. This distribution contains many close PBHs pairs that decouple from the expansion of the Universe before the matter-radiation equality and form an eccentric binary. The angular momentum of the binary arises from the tidal torque generated by the surrounding matter inhomogeneities. The initial population will be affected by interactions with other PBHs in small DM haloes when the PBH fraction is large enough~\cite{Raidal:2018bbj,Vaskonen:2019jpv,Jedamzik:2020ypm,DeLuca:2020jug,Young:2020scc}.\footnote{The effect of distant encounters of the binaries with single PBHs in large haloes was found to be negligible in Ref.~\cite{Young:2020scc}.}. Also accretion can affect the PBH merger rate and generate spins for PBH~\cite{DeLuca:2020bjf,DeLuca:2020fpg,DeLuca:2020qqa}. However, as discussed in the introduction, due to large uncertainties related to accretion modeling, especially in binary systems, we neglect the PBH spins and the accretion effects.

\begin{figure}
\centering
\includegraphics[width=0.56\textwidth]{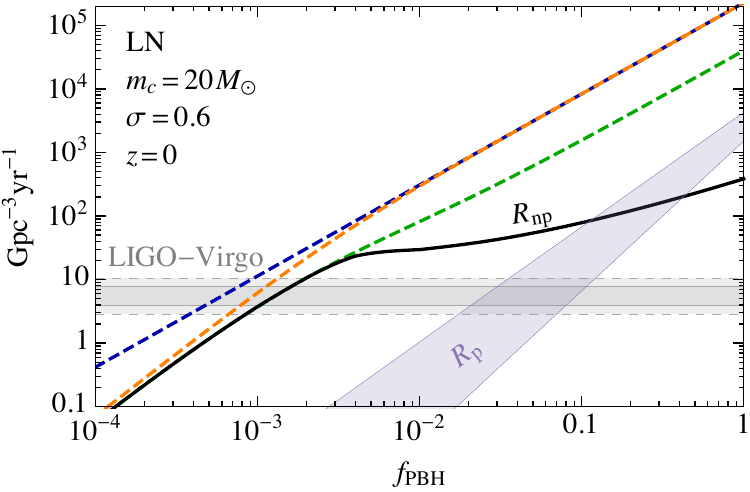}
\caption{The black solid line shows the present merger rate of PBH binaries~\eqref{eq:Rnp} for a log-normal PBH mass function at $m_c = 20\,\Msun$ of width $\sigma = 0.6$. The blue dashed line corresponds to Eq.~\eqref{eq:Rnp} with $S=1$, the orange line includes the suppression arising from the effect of the smooth matter component, the green dashed line from the closest PBH falling into the binary, and the black solid line from the binary being absorbed by a PBH cluster that collapsed before $z=0$. The purple shaded region indicates the amplitude of the merger rate of perturbed PBH binaries. The grey shaded region shows the range $4.9<N_{\rm O2}<17.8$ for O1\&O2 (dashed) and $25.3<N_{\rm O3}<49.4$ for O3a (solid), which are $2\sigma$ CLs from LIGO-Virgo observations for the number of BH merger events.
}
\label{fig:Rnp}
\end{figure}

The merger rate of PBH binaries formed in the early Universe is~\cite{Raidal:2018bbj,Vaskonen:2019jpv}
\be \label{eq:Rnp}
	\frac{\td R_{\rm np}}{\td m_1 \td m_2} \approx \frac{1.6 \times 10^{6}}{\Gpc^{3} \yr} \, f_{\rm PBH}^{\frac{53}{37}}  \left(\frac{t}{t_{0}}\right)^{-\frac{34}{37}} \left(\frac{M}{\Msun}\right)^{-\frac{32}{37}} \eta^{-\frac{34}{37}} S[\psi,f_{\rm PBH},M] \frac{\psi(m_1) \psi(m_2)}{\langle m\rangle^2} \,, 
\ee
where $M=m_1+m_2$ denotes the total mass of the binary, $\eta=m_1 m_2/M^2$ is the symmetric mass ratio of the binary, $\psi(m) \equiv n_{\rm PBH}^{-1} \td n_{\rm PBH}/\td \ln m$ is the PBH mass function,\footnote{We note that in Ref.~\cite{Raidal:2018bbj} the mass function was defined as $\psi(m) \equiv \rho_{\rm PBH}^{-1}\td n_{\rm PBH}/\td \ln m$. With the definition chosen here $\int \psi(m) \td \ln m = 1$ and the average over PBH masses of any quantity $X$ is $\langle X \rangle \equiv n_{\rm PBH}^{-1} \int X \td n_{\rm PBH} = \int X \psi(m) \td \ln m$.} and $S[\psi,f_{\rm PBH}, M]$ is a suppression factor accounting for the effect of the surrounding smooth matter component (not PBHs) on the binary formation and the disruption of the binary by other PBHs. The disruption consists of two effects: First, PBHs close to the initial PBH pair may fall into the binary. Second, if the PBH binary is absorbed by a PBH cluster that disrupts it. The second effect has been explicitly demonstrated in numerical simulations of PBH binary evolution in the early Universe, that is, in small clusters~\cite{Raidal:2018bbj} and in haloes containing up to 1300 PBH~\cite{Jedamzik:2020ypm}. All these effects are shown separately in Fig.~\ref{fig:Rnp}. Based on the cause of disruption, we divide the merger rate suppression factor into two components, $S(z) = S_1 S_2(z)$. The first factor $S_1$ arises from requirements on the initial configuration and is redshift independent, while $S_2(z)$ depends on the evolution of structures in the Universe and therefore depends on redshift.

Assuming that the initial spatial distribution of PBHs was Poisson, the suppression factor $S_1$ is approximated by (see Appendix~\ref{app:S1})
\be\label{eq:S1_appr}
    S_1 \approx 1.42 \left[\frac{\langle m^2\rangle/\langle m \rangle^2}{\bar{N}(y) + C} + \frac{\sigma_{\rm M}^2}{f_{\rm PBH}^2}\right]^{-21/74} e^{-\bar{N}(y)} \,,
\ee
where $\sigma_{\rm M}\simeq 0.004$ is the rescaled variance of matter density perturbations (not PBHs) at the time the binary was formed, $C(f_{\rm PBH})$ is a fitting function given in Appendix~\ref{app:S1}, and $\bar{N}(y)$ is the expected number of PBHs within a sphere of comoving radius $y$ around the initial PBH pair. The minimal distance $y$ of the third PBH can be estimated as~\cite{Raidal:2018bbj}
\be\label{eq:Ny}
    \bar{N}(y) \approx \frac{M}{\langle m \rangle}\frac{f_{\rm PBH}}{f_{\rm PBH} + \sigma_{\rm M}}.
\ee 
The suppression factor $S_1$ accounts for perturbations in the surrounding smooth matter component and discards initial configurations that contain a third PBH within a distance smaller $y$. When $f_{\rm PBH}\lesssim \sigma_{\rm M}$ the suppression factors is dominantly due to inhomogeneities in the smooth matter component, as in that case, the angular momentum of the initial binary is determined by the tidal torque from this matter component. The orange dashed line in Fig.~\ref{fig:Rnp} shows the corresponding merger rate, given by $S_1$ in the limit $\bar{N}(y) \to 0$, that is, omitting restrictions on the distance of the nearest PBH.

To estimate the factor due to PBH-binary scattering within DM structures, we assume that all binaries within haloes of size $N \leq N_c(z)$ are likely to collide at least once with other PBH before redshift $z$ and do therefore not contribute to the merger rate due to their ionization or significant reduction of their eccentricity. In this case, the suppression factor at redshift $z$ is given by the fraction of PBHs outside of haloes or subhaloes containing less than $N_c$ PBH at redshift $z$. It can be estimated as~\cite{Vaskonen:2019jpv} 
\be
    S_2(z) = 1 - \sum_{N=3}^{N_{c}(z)} \left[p_{N}(z_c) + \sum_{N'>N_{c}(z)} p_{N,N'}(z_c) \right] \,,
\ee
assuming a relatively narrow PBH mass function. Above, $z_c = f_{\rm PBH} z_{\rm eq}/\sqrt{N_c}$, with $z_{\rm eq}$ the redshift of matter-radiation equality, denotes the redshift at which the haloes of $N_c$ PBHs form, and
\bea
& p_{N}(z) = \frac{N^{-1/2}e^{-N/N^{*}(z)}}{\sum_{N>2} N^{-1/2}e^{-N/N^{*}(z)}}\,, \qquad
& p_{N,N'}(z) = p_{N'}(z) \,\frac{N^{-1/2}e^{-N/N^{*}(z)}}{\sum_{N = 2}^{N'} N^{-1/2}e^{-N/N^{*}(z)}} ,
\eea
describe, respectively, the probability of the binary being part of a halo containing $N$ PBHs at redshift $z$ and the probability of the binary being part of a subhalo containing $N$ PBHs inside a halo of $N'$ PBHs. $N^*(z)$ is the characteristic number of PBHs in a halo at redshift $z$, and can be estimated by~\cite{Inman:2019wvr}
\bea
&    N^*(z) = \left[\ln(1+\delta^*) - \delta^*/(1+\delta^*)\right]^{-1} , \\
&    \delta^* \approx 1.69 \left[f_{\rm PBH}\, {}_1F{}_2\left((1-\sqrt{21})/4,(1+\sqrt{21})/4;1;-(z_{\rm eq}+1)/(z+1) \right) \right]^{-1}\,.
\eea
Haloes (or subhaloes) containing $N>N_c(z)$ PBHs, collisions between PBH are unlikely to take place at redshifts larger than $z$. The critical number $N_c(z)$ could be chosen based on the characteristic timescale for 2-body collisions, the timescale for evaporation or of gravitational instability manifested through core collapse. The last option is the most conservative as it corresponds to the largest $N_c(z)$ and thus the smallest merger rate. It is given by the solution to $18 t_r = t(z)$, where $t_r \approx 2\,{\rm kyr}\,N_c^{7/4}f_{\rm PBH}^{-5/2} /\ln (N_c/f_{\rm PBH})$ is the relaxation time of a halo containing $N$ BHs~\cite{Quinlan:1996bw}.  By numerical fit we find that at $z=0$ the suppression factor $S_2(z)$ can be approximated as 
\be \label{eq:S2a}
    S_2(z=0) \approx \min\left[1,\,9.6\times 10^{-3} f_{\rm PBH}^{-0.65} e^{0.03 \ln^2f_{\rm PBH}}\right]\,,
\ee
and a good approximation at $z\lsim 100$ is obtained by replacement $f_{\rm PBH} \to (t(z)/t_0)^{0.44} f_{\rm PBH}$ in~\eqref{eq:S2a}. As seen from Fig.~\ref{fig:Rnp}, for $M = 20 \Msun$ this suppression factor starts to be relevant when $f_{\rm PBH} \gsim 0.003$.

In Fig.~\ref{fig:Rnp} we also show an estimate for the merger rate $R_p$ of binaries that are disrupted. This can be greater than the merger rate of non-perturbed binaries $R_{\rm np}$ if almost all DM consists of PBHs. An important feature of this component is, that perturbed primordial binaries that merge within the Hubble time are much harder, and thus unlikely to be ionized, and, unlike perturbed ones, not very eccentric (for details, see~\cite{Vaskonen:2019jpv}.) As a result, binary-PBH collisions will not significantly affect the merger rate of this population and thus $R_p$ can be used to constrain $f_{\rm PBH} > 0.1$ in scenarios where PBH collisions within the DM substructure are likely, such as for initially clustered PBHs~\cite{Clesse:2016vqa,Garcia-Bellido:2017xvr,Trashorras:2020mwn}. 

We remark that the PBH merger rate in scenarios with an initially non-Poisson distribution depends on the specifics of how the deviation from a Poisson distribution is realized. In particular, if the PBHs are extremely strongly clustered so that most of the initial binaries merge before today, it might be possible to evade the merger rate constraints from early binaries~\cite{Raidal:2017mfl,Atal:2020igj}. In such scenarios, the initial PBH binaries merging today tend to be almost circular and hard. Thus their coalescence time is not strongly affected by collisions with other PBHs. However, strong clustering implies a peak in the power spectrum a few orders of magnitude below the scale at which the PBHs are produced~\cite{Atal:2020igj}. Thus, for PBHs in the LIGO-Virgo mass range clustering of PBH when $f_{\rm PBH} \approx 1$ might be potentially probed by observing the $\mu$-distortion of CMB. Also, in case the BH binary coalescence time is significantly shortened by clustering, the merging PBH will produce a stochastic gravitational wave background (SGWB) that might be observable by LIGO-Virgo even if the current merger rate is negligible. The produced SGWB depends on the specifics of how the redshift dependence of the merger rate is modified when compared to the Poisson-distributed case (see, \eg,~\cite{Raidal:2017mfl,Atal:2020igj}).


Finally, going back to the accretion effects on the merger rate, in Refs.~\cite{DeLuca:2020qqa,Wong:2020yig}, the different accretion scenarios for PBHs were parametrized by a cut-off redshift $z_{\rm cut-off}$ below which accretion contributes negligibly to the evolution of the binary. A recent Bayesian analysis of the GWTC-2 dataset~\cite{Wong:2020yig} found that, when omitting spin data, the observed mergers prefer PBH models with $z_{\rm cut-off} \gtrsim 25$ at $1\sigma$ CL implying that the observed mass and redshift distributions are consistent with PBH models in which accretion effects are negligible. However, fits involving spin data find $z_{\rm cut-off} \approx 20\pm 3$ thus favouring PBH models with non-negligible accretion. This is intuitively clear as accretion onto PBHs is required to accumulate sufficient angular momentum. Therefore, given the spin-independent analysis of Ref.~\cite{Wong:2020yig}, we expect that the omission of accretion effects will not significantly affect our conclusions.

\section{PBHs in the light LIGO-Virgo observations}
\label{sec:implications}


PBH binaries can produce GW signals detectable with current or planned GW observatories. The loudest events can be detected individually, while the weakest ones will contribute to the SGWB. If $\mathcal{O}(10\Msun)$ PBHs comprise a large fraction of DM, the present PBH merger rate is predicted to exceed the range indicated by the LIGO-Virgo observations~\cite{LIGOScientific:2018mvr,Abbott:2020niy} as can be seen from Fig.~\ref{fig:Rnp} and therefore their abundance can be constrained by the LIGO-Virgo observations~\cite{Raidal:2017mfl,Ali-Haimoud:2017rtz,Raidal:2018bbj,Kavanagh:2018ggo,Authors:2019qbw,Vaskonen:2019jpv,DeLuca:2020qqa}. In the following, we will update these constraints, including the LIGO-Virgo O3a results. In addition, as in Refs.~\cite{Raidal:2017mfl,Raidal:2018bbj,Chen:2018czv,Gow:2019pok,Wu:2020drm,Hall:2020daa,DeLuca:2020qqa,Wong:2020yig}, we also consider the possibility that at least some of the observed GW events are from PBH mergers, and perform fits of the PBH abundance and mass function including the latest LIGO-Virgo observations.

\subsection{Likelihood analysis}

We select from the GWTC-1 and GWTC-2 catalogues~\cite{LIGOScientific:2018mvr,Abbott:2020niy} the events for which both of the components are heavier than $3\Msun$, to exclude binaries containing potential neutron stars. With this criterion, during O1 and O2 LIGO-Virgo network observed 10 BH-BH binaries. Assuming Poisson statistics, the $2\sigma$ CL range for the expected number of events is $4.9<N_{\rm O2}<17.8$. During O3a additional 36 BH-BH binaries were observed, and the corresponding $2\sigma$ CL range is $25.3<N_{\rm O3}<49.4$. The corresponding merger rate is show by the grey lines in Fig.~\ref{fig:Rnp} assuming that the merger rate is constant and the PBH follow a log-normal BH mass function (see below) with $m_c = 20\Msun$ and $\sigma = 0.6$.

To study the characteristics of the PBH population needed to explain the LIGO-Virgo observations, we perform a likelihood fit for the PBH mass function and abundance. Observing independent events characterized by the GW data $d_{j}$ can be modelled as an inhomogeneous Poisson process. The corresponding likelihood is\footnote{The Poisson factor $\propto N^{N_{\rm obs}}e^{-N}$ in the likelihood was not included in earlier analyses of Refs.~\cite{Raidal:2018bbj,DeLuca:2020qqa}. Moreover, we do not include the $\theta({\rm SNR} - {\rm SNR}_c)$ factor for rejecting events with a low signal-to-noise-ratio as the likelihood is constructed from observed events~\cite{Mandel:2018mve}. We also omit the integral over the observation time as it would contribute only by a constant factor. So the observation time and the detector sensitivity enter only through $N$.}
\be \label{eq:lpsi}
    \ell \propto e^{-N}\prod_{j=1}^{N_{\rm obs}} \int p_j(d_{j}|m_1,m_2,z) \,\td \lambda  \,,
\ee
where $p_j(d_{j}|m_1,m_2,z)$ is the source likelihood, $\lambda$ is the expected number of GW signals from BH binary mergers arriving per unit time and $N$ is the expected number of observed events within a time interval. The expected number of GW signals from BH binary mergers with progenitor masses $m_1$ and $m_2$ at redshift $z$ is given by the merger rate $R$ as
\be\label{eq:Nevents}
    \td \lambda = \frac{1}{1+z}\frac{\td R}{\td m_1 \td m_2} \frac{\td V_c}{\td z}  \td m_1 \td m_2 \td z \,,
\ee
where $V_c$ is the comoving Hubble horizon volume. Integrating over $\td \lambda$ and accounting for the detection probability $p_{\rm det}$, we get the expected number of observed events within a time interval $\mathcal{T}$,
\be \label{eq:N}
    N \equiv  \int_0^{\mathcal{T}} \td t \, \int \td \lambda \, p_{\rm det}({\rm SNR}_c/{\rm SNR}) \,,
\ee
where the detection probability estimate used here is given in appendix~\ref{app:detdectability}. We approximate the source likelihood as
\be
    p_j(d_{j}|m_1,m_2,z) \approx p_j(m_{j,1} | m_1) p_j(m_{j,2} | m_2) p_j(z_j | z) \,.
\ee
The functions $p_j(m_j | m)$ and $p_j(z_j | z)$ account for the experimental uncertainties in the measurement of the BH masses $m_{j,1}$ and $m_{j,2}$ and redshifts $z_j$, which we approximate with piecewise Gaussian distributions,
\be
p_j(x_j|x) = \sqrt{\frac{2}{\pi}} \,(\sigma_{j,1} + \sigma_{j,2})^{-1}\, 
\begin{cases}
e^{-(x-x_j)^2/(2\sigma_{j,1}^2)} & x<x_j \,,\\
e^{-(x-x_j)^2/(2\sigma_{j,2}^2)} & x\geq x_j \,,
\end{cases}
\ee
with the mean and variances taken from the GWTC-1 and GWTC-2 catalogues~\cite{LIGOScientific:2018mvr,Abbott:2020niy}.~\footnote{We note that the errors in the catalogues are given at 90\% CL.} We assume standard cosmology~\cite{Ade:2015xua}, neglect potential correlations between progenitor masses and redshifts, and take the merger rate to be independent of the spins of the BHs.~\footnote{By neglecting the correlations we slightly a overestimate of the uncertainties.} For a given PBH mass function $\psi$ we can then perform a maximum likelihood analysis.

\begin{figure}
\centering
\includegraphics[height=0.25\textwidth]{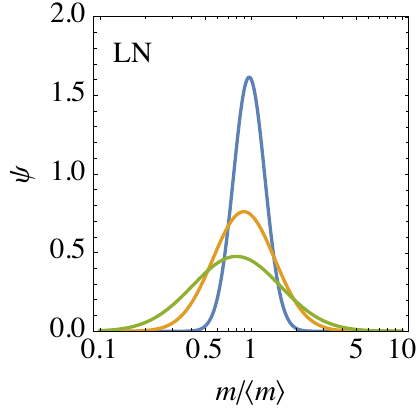}
\includegraphics[height=0.25\textwidth]{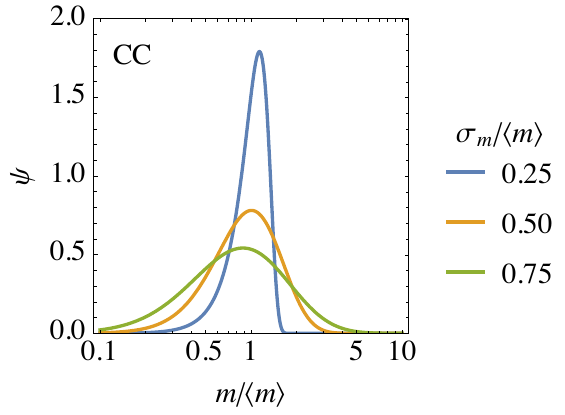} \hspace{1mm}
\includegraphics[height=0.25\textwidth]{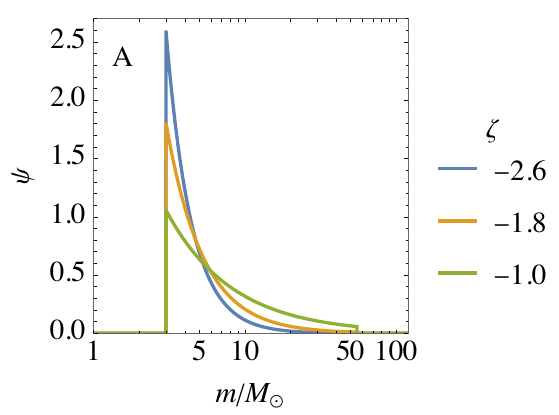}
\caption{BH mass functions used in this work. The left and middle panels show the log-normal and critical collapse PBH mass functions. The right panel shows the truncated power-law mass function used for modelling the astrophysical background.}
\label{fig:psi}
\end{figure}

We consider two PBH mass functions shown in the left and middle panels of Fig.~\ref{fig:psi}:
\begin{enumerate}
\item Log-normal mass function~\cite{Dolgov:1992pu}:
\be
    \psi(m) = \frac{1}{\sqrt{2\pi} \sigma} \exp\left[-\frac{\ln^2(m/m_c)}{2\sigma^2}\right] \,,
\ee
where, in the log-scale, $m_c$ is the mean mass and $\sigma$ the width of the distribution. The mean mass and variance of the log-normal mass function are 
\be
    \langle m\rangle = m_c e^{\sigma^2/2} \,, \quad 
    \sigma_m^2 = \langle m\rangle^2 (e^{\sigma^2} - 1) \,.
\ee
In many studies this mass function has been used as a benchmark model for PBHs (see, \eg,  Refs.~\cite{Carr:2017jsz,Raidal:2017mfl,DeLuca:2020qqa}), and provides a good approximation in various PBH formation scenarios (see, \eg,  Refs.~\cite{Green:2016xgy,Kannike:2017bxn}) if the critical scaling relationship~\cite{Choptuik:1992jv,Niemeyer:1999ak,Musco:2008hv} is not accounted for in the gravitational collapse.

\item Critical collapse mass function~\cite{Niemeyer:1997mt,Byrnes:2018clq,Vaskonen:2020lbd}:
\be
    \psi(m) = c_0 m^{1+1/\gamma} \exp\left[-c_1 (m/\langle m\rangle)^{c_2}\right] \,,
\ee
where $\gamma = 0.36$ is a universal exponent related to the critical collapse of radiation~\cite{Choptuik:1992jv,Evans:1994pj}, and the factors $c_0$ and $c_1$ are chosen such that $\int \psi(m) \td\ln m = 1$ and $\int \psi(m) \td m = \langle m\rangle$. The variance of the critical collapse mass function is 
\be
    \sigma_m^2 
    = \langle m\rangle^2 \left[\frac{\Gamma\left((1+\gamma^{-1})/c_2\right) \Gamma\left((3+\gamma^{-1})/c_2\right)}{\Gamma\left((2+\gamma^{-1})/c_2\right)}-1\right] \,.
\ee
This mass function is obtained when the critical scaling relation~\cite{Choptuik:1992jv,Niemeyer:1999ak,Musco:2008hv} is accounted for in PBH formation.
\end{enumerate}

\begin{figure}
\centering
\includegraphics[width=0.98\textwidth]{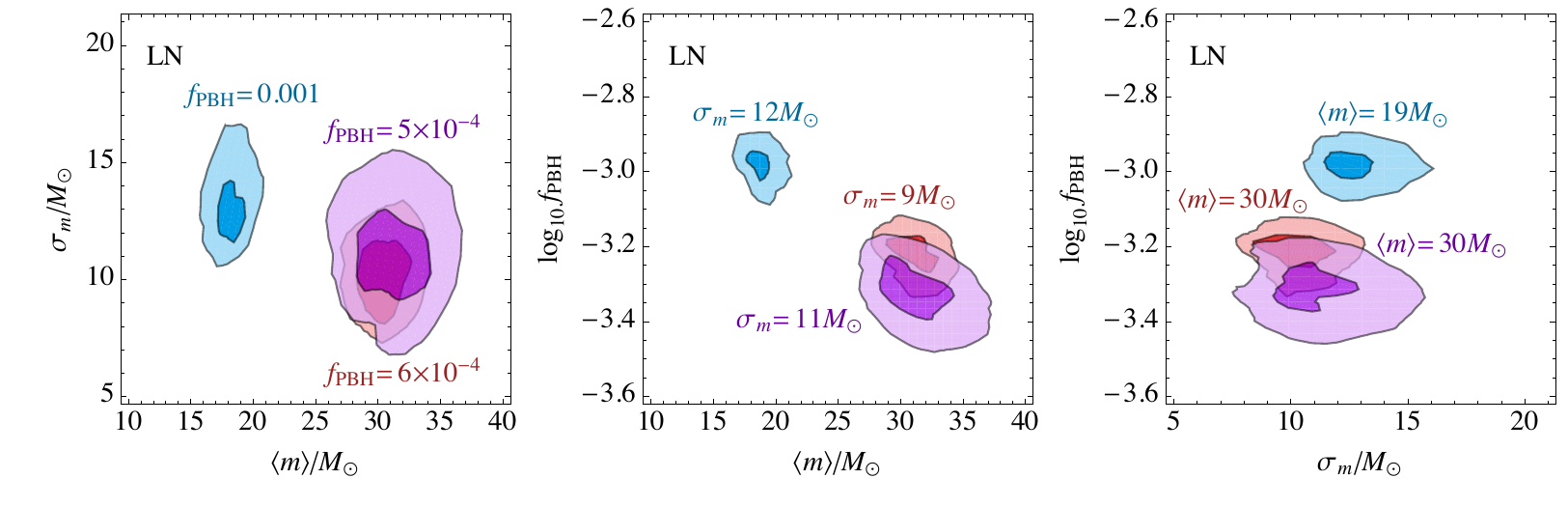}
\includegraphics[width=0.98\textwidth]{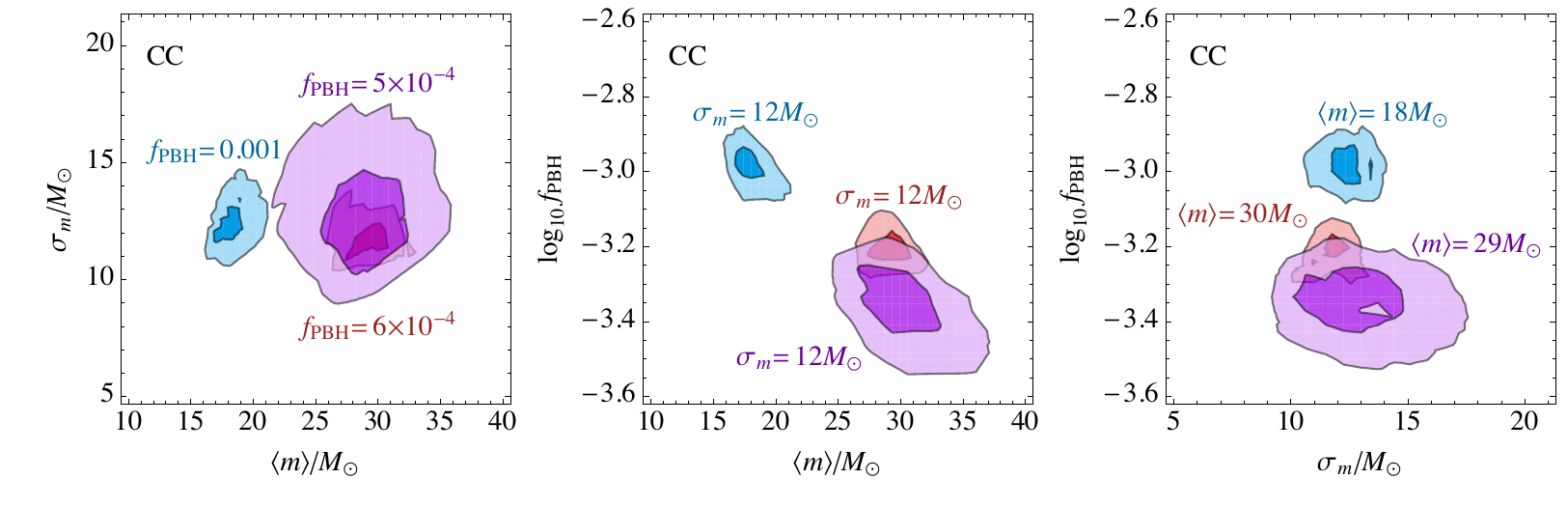}
\caption{The shaded regions show the $1\sigma$ and $2\sigma$ CL regions from maximum likelihood fits of the PBH merger rate on all LIGO-Virgo events (blue) and on the selected events (red). The purple regions show the result of the combined fit of primordial and astrophysical BH merger rates. The upper panels show the fits for the log-normal PBH mass function and the lower panels for the critical collapse PBH mass function. In each case, three cuts of the likelihood function through the best-fit point of each fit are shown, with the value of the third parameter indicated for each fit with the corresponding colors in the plots}.
\label{fig:Pfit}
\end{figure}

The results of the maximum likelihood analysis, obtained by randomly scanning the parameter space, for the log-normal and critical collapse mass functions are shown in Fig.~\ref{fig:Pfit} by the blue regions, assuming that all observed events are from PBH binaries formed in the early Universe. We see that the fits for these mass functions are very similar, indicating that the current data does not differentiate well between the different peak shapes as long as their position and width are fixed.

\begin{figure}
\centering
\includegraphics[width=0.9\textwidth]{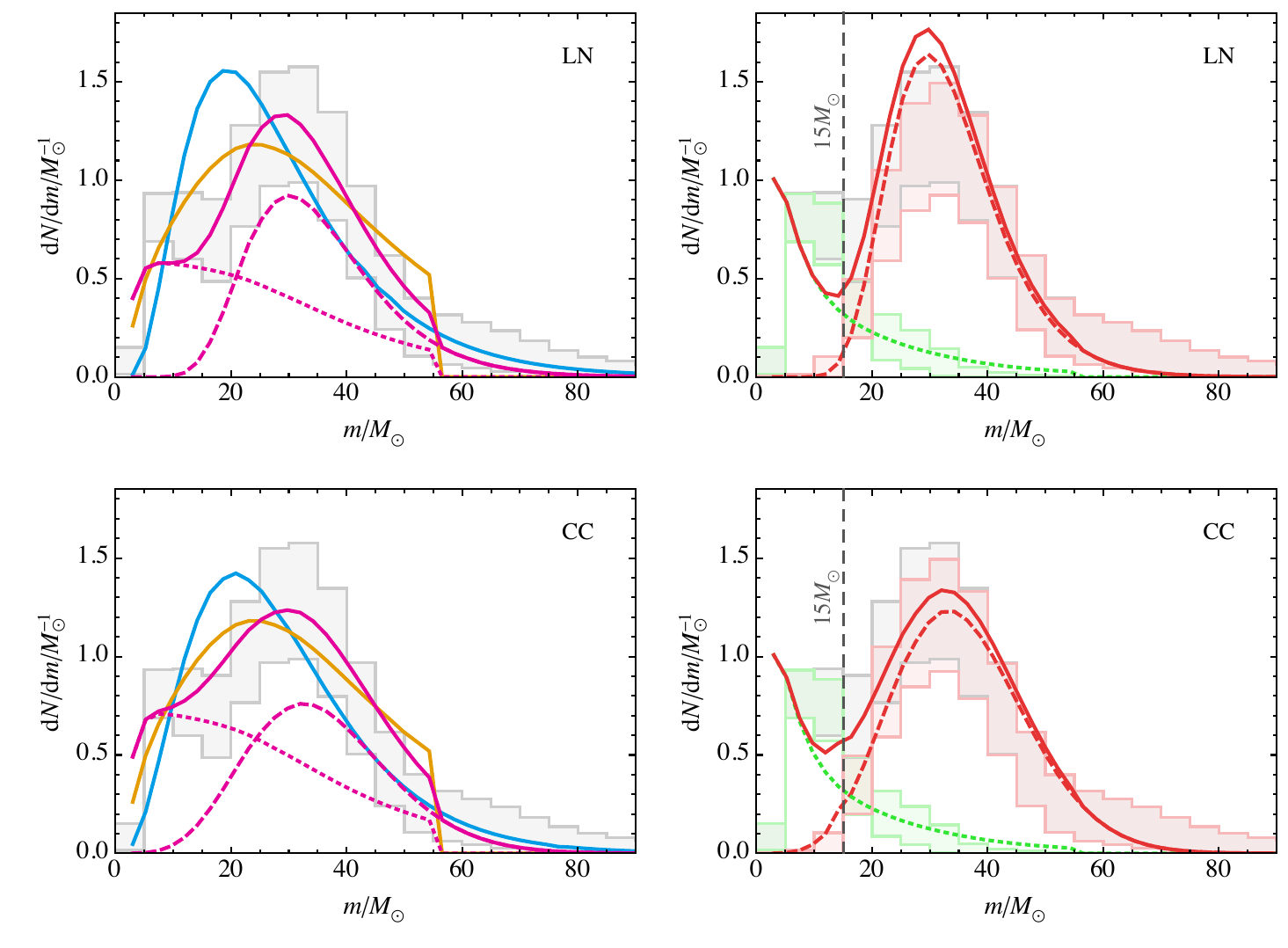}
\caption{
The best-fit mass distributions assuming a log-normal PBH mass function (upper panels) and a critical collapse PBH mass function (lower panels). \emph{Left panels:} The blue and yellow curves correspond, respectively, to the PBH and ABH model fits to all observed BH merger events. The purple curve instead corresponds to the fit combining the ABH and PBH models. The purple dotted and dashed curves indicate, respectively, the contributions from PBHs and ABHs. The grey band shows the binned observed population of BH masses. The uncertainty of the number of BH in each bin corresponds to a $1\sigma$ CL and is inferred from the uncertainties in the BH masses only. Correlations between progenitor masses are ignored. \emph{Right panels:} The green and red dashed curves correspond, respectively, to the ABH fit to the events containing at least one BH lighter than $15\Msun$, and the PBH fit to the remaining events. The green and red histograms depict the corresponding observed mass distributions in the binary sub-populations. The red solid curve shows the sum of the PBH and ABH mass distributions.}
\label{fig:bestfitMF}
\end{figure}

To consider a scenario where only a fraction of the events has a primordial origin, we split the events into two subsets by hand. The two subsets are constructed so that the first subset contains all events with at least one of the progenitors lighter than $15\Msun$ while the second subset consists of the rest of the events.\footnote{The separation scale $15\Msun$ is chosen as the approximate beginning of an apparent peaked feature in the histogram.}. This is illustrated in the right panels of Fig.~\ref{fig:bestfitMF}. The red regions in Fig.~\ref{fig:Pfit} show the fits of the PBH models to only the events belonging to the second subset. As expected, in this case, the fit prefers mass functions that peak at slightly higher masses and are narrower than in the fit to all observed events. Again, no significant differences in the fits for the log-normal and critical collapse mass functions are found.

\subsection{Constraints}

\begin{figure}
\centering
\includegraphics[width=0.45\textwidth]{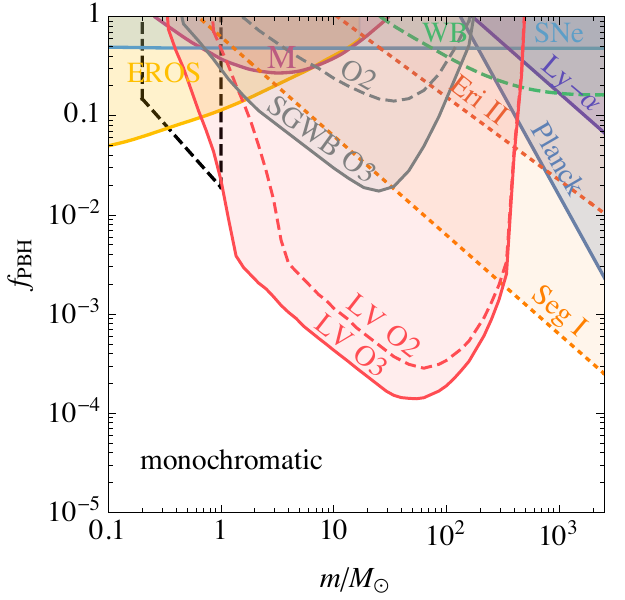} \hspace{2mm}
\includegraphics[width=0.45\textwidth]{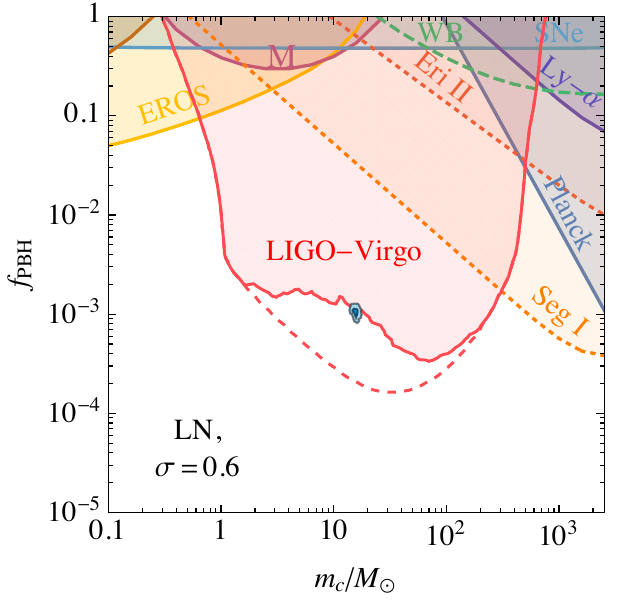}
\caption{\emph{Left panel:} Constraints for monochromatic PBH mass function. The solid and dashed red lines show the $2\sigma$ CL constraint from LIGO-Virgo (LV) O3a obtained assuming that all observed events are astrophysical and the grey solid and dashed lines show the constraints from the non-observation of the SGWB by LIGO-Virgo. The red and grey dashed curves instead depict the corresponding O2 constraints. \emph{Right panel:} Constraints for log-normal PBH mass function with $\sigma = 0.6$. The red dashed curve shows the $2\sigma$ CL constraint from LIGO-Virgo obtained assuming that all observed events are astrophysical and the solid red curve presents the $2\sigma$ CL constraint when the observed events are taken into account. The blue region right below the solid red curve indicates the PBH fit to all observed events.}
\label{fig:constr}
\end{figure}

Assuming that all events observed by the LIGO-Virgo network are astrophysical, 95\% CL upper bound on the PBH abundance is obtained by requiring that $N<3$. In the left panel of Fig.~\ref{fig:constr} shows this bound for a monochromatic PBH mass function by the red dashed and solid lines for O1\&O2 and O3a. The constraints from non-observation of the SGWB from PBH binaries from O1\&O2 and O3a are instead shown by the grey dashed and solid lines, respectively. The constraint from non-observation of the SGWB is obtained by requiring that the signal-to-noise ratio of the SGWB for LIGO-Virgo detectors is smaller than a threshold value, ${\rm SNR}_{\rm BG} < {\rm SNR}_c$ (see Appendix~\ref{app:detdectability} for details).

We emphasize that, whereas the merger rate constraint depends on whether some or all of the observed events are assumed to be primordial, the constraint from non-observation of the SGWB is independent of the origin of LIGO-Virgo events. The bound from non-observation of the SGWB has gotten almost an order of magnitude stronger when compared to the O1\&O2 bound, whereas the improvement in the $N<3$ -bound is by a factor of $3$.

In the right panel of Fig.~\ref{fig:constr} we show the LIGO-Virgo constraints assuming a log-normal PBH mass function with $\sigma = 0.6$ (corresponding to its best-fit value). The red dashed curve shows the $N<3$ constraint, which corresponds to the $2\sigma$ CL constraint obtained from the likelihood $\ell \propto e^{-N}$, that is, to not fitting any of the observed events.

It is possible that a subset of the events has a primordial origin, while the rest of the events are astrophysical. Different ways of splitting the data into primordial and astrophysical events will provide different constraints on $f_{\rm PBH}$, all of which will be weaker than the $N<3$ constraint by construction. Moreover, the case where all events are primordial does not necessarily give the weakest constraint -- even if the number of primordial events in any proper subset is smaller, the corresponding constraint can be weaker in some mass range. For example, the current data disfavours binary populations with a narrow mass function centered around 40 $\Msun$, thus a subset containing events with progenitor masses around 40 $\Msun$ might provide a larger likelihood and thus a weaker constraint on $f_{\rm PBH}$ around this mass than the full dataset. Therefore, in order to obtain a more conservative constraint, it is necessary to consider different splittings of the data into subsets of primordial and non-primordial events. With 46 events, there are $2^{46}$ ways of splitting the data. Computing the likelihood for all of them is not feasible as there are $2^{46}$ possible subsets. However, we find that the constraint converges quite fast as more subsets are included. We exclude events with mixed origin, i.e. assume that both progenitors of each event are either astrophysical or primordial.

In the right panel of Fig.~\ref{fig:constr}, we show by the solid red line the constraint obtained by combining the likelihoods for 270 different subsets of the observed events as $\max_j[\ell_j/\max \ell_j]$, where the index $j$ refers to different subsets. The red shaded region is therefore excluded independently of whether some of the observed events are PBH mergers or not. For comparison, the right panel of Fig.~\ref{fig:constr} depicts the fit assuming that all observed events are primordial. As expected, the $2\sigma$ contour of the fit touches the $2\sigma$ upper bound on $f_{\rm PBH}$. By construction, any point in the region between the red and red dashed lines corresponds to at least one scenario in which some of the progenitors of the LIGO-Virgo GW events were primordial.

Other shaded regions in Fig.~\ref{fig:constr} show the constraints from microlensing results from EROS~\cite{Tisserand:2006zx}, OGLE~\cite{Niikura:2019kqi}, and MACHO~\cite{Allsman:2000kg}, lensing of type Ia supernovae (SNe)~\cite{Zumalacarregui:2017qqd}, survival of a stars in dwarf galaxies (Seg~I and Eri~II)~\cite{Brandt:2016aco,Koushiappas:2017chw}, distribution of wide binaries (WB)~\cite{Monroy-Rodriguez:2014ula}, Lyman-$\alpha$ forest data~\cite{Afshordi:2003zb,Murgia:2019duy} and limits on accretion (Planck)~\cite{Ricotti:2007au,Horowitz:2016lib,Ali-Haimoud:2016mbv,Poulin:2017bwe,Hektor:2018qqw,Serpico:2020ehh}, and they are converted to the log-normal mass function using the method introduced in Ref.~\cite{Carr:2017jsz}. We see that the LIGO-Virgo observations give the strongest constraint in the mass range $0.6-400\Msun$. 

The left panel of Fig.~\ref{fig:constr} includes the constraint for subsolar mass PBH from the LIGO-Virgo O2 run~\cite{Abbott:2018oah,Authors:2019qbw}. It roughly matches the tail of our LIGO-Virgo O2 constraint if we omitted the suppression factors described in Sec.~\ref{sec:mergers}. However, accounting for the disruption of binaries eliminates the O2 constraint for subsolar mass PBHs almost completely. Nevertheless, the O3 constraint is able to compete with constraints from lensing in the subsolar mass range.

\section{Astrophysical vs. primordial BH binaries}
\label{sec:discussion}

\subsection{Astrophysical BH binaries}

In astrophysics typically two main pathways leading to tight compact object binaries are discussed: (i) isolated (or 'field') binary evolution, (ii) dynamical capture in dense stellar environments. Other possible channels are also discussed in literature, \eg, mergers in AGN gas disks or mergers in triple systems assisted by the Kozai-Lidov mechanism. For a recent review see Ref.~\cite{Barack:2018yly}. 

The majority of massive stars ($>70\%$) are naturally born as binary systems. In the usual scenario, a common envelope stage of late time evolution is crucial for tightening the binary before the gravitational wave losses can become efficient, enabling the system to collide within the Hubble time. However, the common envelope phase involves significant uncertainties~\cite{Ivanova:2012vx}, making the predictions highly unreliable. There is a possibility to bypass the complexities of the common envelope phase by starting with an initial close binary with both components having enough rotation. This drives mixing currents inside the stars and results in chemically homogeneous evolution~\cite{Marchant:2016wow,deMink:2016vkw,Eldridge:2016ymr,Woosley:2016nnw}. In this case, the stars do not bulge up during the later stages of evolution, allowing them to be initially much closer compared to the usual case. The massive compact stellar remnants form in supernova explosions, which, due to asymmetries, typically produce a level of kick-back that might easily disrupt the binary system. Due to these uncertainties the estimated rates of BH-BH mergers produced via isolated binary evolution vary quite substantially: $10-300\,\Gpc^{-3}\yr^{-1}$~\cite{Eldridge:2016ymr,Belczynski:2016obo,Stevenson:2017tfq,Kruckow:2018slo}.

Regarding the component spins, assuming these are large enough to be measurable, the binary evolution makes a prediction that these should be preferentially aligned with the system's orbital angular momentum vector. Despite supernova kicks, which will surely introduce a level of smearing, the predicted spin-orbit alignment is one of the main distinguishing characteristics for this channel.

The latest LIGO-Virgo data, which have some statistical sensitivity to spins, demonstrate that not all of the events can be produced via this `spin-aligned channel'~\cite{Abbott:2020gyp}. Also in this channel, there are potential problems to produce the most massive merger observed thus far -- GW190521~\cite{Abbott:2020tfl}. Here at least one of the BH masses falls into the so-called pair-instability supernova (PISN) gap~\cite{1967ApJ...148..803R,1968Ap&SS...2...96F,2002RvMP...74.1015W} -- standard stellar evolution should not produce BHs of this mass as these should be destroyed by the resulting PISN explosion. The PBH scenario for GW190521 was discussed in Ref.~\cite{DeLuca:2020sae}.

The PISN mass gap is expected to fall in the range $\sim 50-130\,\Msun$. The progenitors of PISN are low to moderate metallicity stars with initial masses in the range $\sim 130-250\,\Msun$~\cite{2002RvMP...74.1015W,2003ApJ...591..288H}. BHs below the lower edge of the mass gap are due to progenitors with somewhat lower initial masses, $\sim 100-130\,\Msun$. These stars produce the so-called pulsational PISN (PPISN)~\cite{2002RvMP...74.1015W,2002ApJ...567..532H,2003ApJ...591..288H}, where the star after several violent pulsational mass loss events leaves behind remnant BH. LIGO-Virgo collaboration in its template-based analysis, for cases where the template allows for peaks, pick up a specific mass scale $\sim 30-40\,\Msun$~\cite{2002RvMP...74.1015W}, which is expected as a pile-up region produced by the progenitor PPISN events.


Some of the above tensions could be relaxed by the other -- dynamical capture channel. Dynamical capture can be effective only in very dense stellar environments like globular clusters (GCs) and nuclear star clusters. Since the relative fraction of stars in these dense regions is low compared to the field, the predicted merger rates are also typically somewhat lower, reaching $\sim 10\,{\rm Gpc}^{-3}{\rm yr}^{-1}$~\cite{Rodriguez:2016kxx,Askar:2016jwt}. This channel predicts BH spins to be uncorrelated, \ie, isotropically distributed with respect to the system's orbital angular momentum, which seems to be mostly consistent with the latest LIGO-Virgo data. In very dense regions and inside deep enough potential wells, the BH-BH collisions can go through more than one generation of mergers, \ie, proceed in a hierarchical fashion~\cite{Rodriguez:2019huv}. This might help to populate the above-mentioned PISN mass gap and explain the occurrence of events such as GW190521~\cite{Kimball:2020qyd}.

At the moment, quite possibly, the merger event that astrophysics struggles the most is GW190814. Although this event involves a compact object with the somewhat unexpected mass of $\sim 2.6\,M_{\odot}$, thus being not clear if it is a light BH or heavy NS, the greatest surprise is its extreme mass ratio $\sim 1:9$. A possible scenario for its formation is discussed in Refs.~\cite{Liu:2020gif,Lu:2020gfh}.

\subsection{Has LIGO-Virgo seen PBH mergers?}

The combination of different astrophysical channels seems to provide a sufficiently flexible framework to account for the observed phenomenology fully; that is, in this case, there is no room for PBH events. This is the motivation behind our most extreme PBH bounds shown in Fig.~\ref{fig:constr}. On the other hand, the simplest PBH models, predicting slowly rotating BHs, can also provide a reasonable match to the observed data. This motivates our PBH fits to the observed events shown in Fig.~\ref{fig:Pfit}. Next, we will include an astrophysical component to the fit and try to answer to the question of whether LIGO-Virgo has seen PBH mergers. We start by introducing a simple model for the merger rate of ABHs.

The redshift dependence of the merger rate is an essential characteristic of the BH binary population~\cite{Fishbach:2018edt}. The PBH merger rate~\eqref{eq:Rnp} has a nearly universal temporal scaling behaviour $R \propto t^{-34/37}$ set by the PBH binary formation mechanism. Similarly, for ABHs, the redshift dependence of the merger rate follows from the star formation rate and the delay from the formation of a stellar binary to merging of the BHs. The star formation rate can be estimated as~\cite{Madau:2016jbv}
\be
	{\rm SFR}(z) \propto \frac{(1 + z)^{2.6}}{1 + ((1 + z)/3.2)^{6.2}} \equiv P_b(z)\,.
\ee
Following Ref.~\cite{Belczynski:2016obo}, we take $P_d(t) \propto t^{-1}$ with $t > 50\,\rm{Myr}$ as the delay time distribution. The differential merger rate of ABH binaries is then
\be \label{eq:RABH}
	\frac{\td R_A}{\td m_1 \td m_2 \td z} = \frac{R_0}{Z_\psi} M^\alpha \eta^\beta \psi(m_1)\psi(m_2) \int \td t_d \td z_b P_b(z_b) P_d(t_d) \delta(t(z)-t(z_b)-t_d) \,,
\ee
where $\alpha$ and $\beta$ parametrize the mass dependence of the astrophysical binary formation, the normalization factor $Z_\psi$ is defined such that $\td R/\td z = R_0$ at $z = 0$. We will restrict our current analysis to the model given above. More general setups were considered in Ref.~\cite{Abbott:2020gyp}, where it was found that assuming a general power-law dependence $R \propto (1+z)^{\kappa}$, a growing rate is favoured. However, the exponent can vary as much as $\kappa = 1.8^{+2.1}_{-2.2}$. Thus it is not yet possible to discriminate between astrophysical and primordial models.

\begin{figure}
\centering
\includegraphics[height=0.38\textwidth]{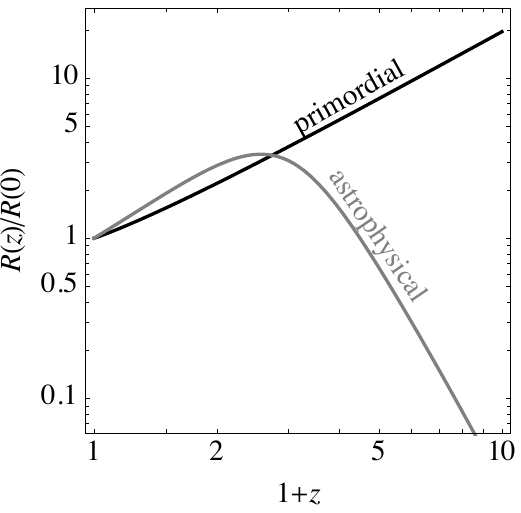} \hspace{5mm}
\includegraphics[height=0.38\textwidth]{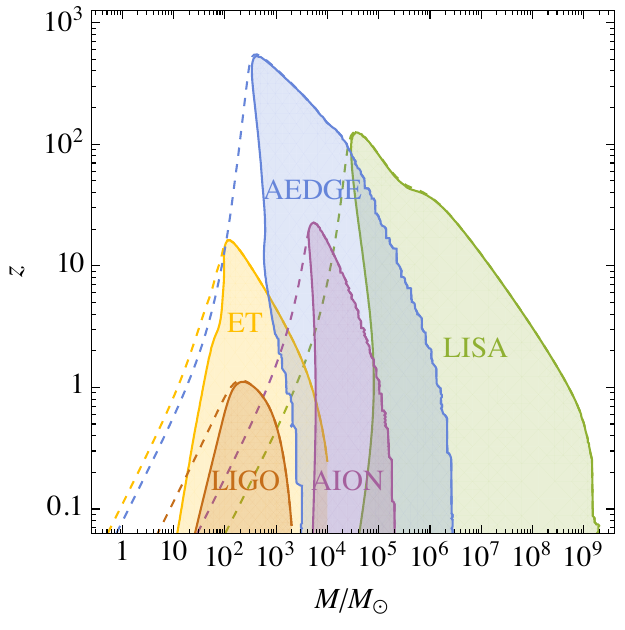}
\caption{\emph{Left panel:} Comparison of the $z$ dependencies of the astrophysical and primordial BH merger rates. \emph{Right panel:} Expected sensitivities of Einstein Telescope (ET)~\cite{Punturo:2010zz}, AION/MAGIS~\cite{Graham:2017pmn,Badurina:2019hst}, AEDGE~\cite{Bertoldi:2019tck}, and LISA~\cite{LISA}, and the design sensitivity of LIGO~\cite{TheLIGOScientific:2014jea} on equal mass BH binaries of total mass $M$ at signal-to-noise ratio larger than 8. In the shaded regions the merger is seen and in the region between the dashed and solid lines only the inspiral phase is seen.}
\label{fig:rates}
\end{figure}

The future instruments with better sensitivity allow us to probe higher redshifts, which is crucial for unambiguously disentangling PBH signal from the astrophysical backgrounds. PBHs, being produced in the early Universe, have significantly more merger activity at high redshifts compared to the ABHs, which, in a broad sense, follow a `filtered version' of the global star formation rate. This is clear from Fig.~\ref{fig:rates} which shows the redshift dependences of the ABH and PBH merger rates: Whereas the PBH merger rate monotonically increases as a function of $z$, the ABH merger rate starts to drop above $z\simeq1$. As shown in the right panel of Fig.~\ref{fig:rates}, LIGO can observe BH mergers only at low redshifts, $z\lsim 1$, but future GW observatories such as ET or AEDGE can instead probe much higher redshifts, $z\simeq10$, for $M=\mathcal{O}(10\Msun)$. Therefore, future GW observations can determine the origin of the BH populations seen by the LIGO-Virgo observatories. 

\begin{figure}
\centering
\includegraphics[height=0.4\textwidth]{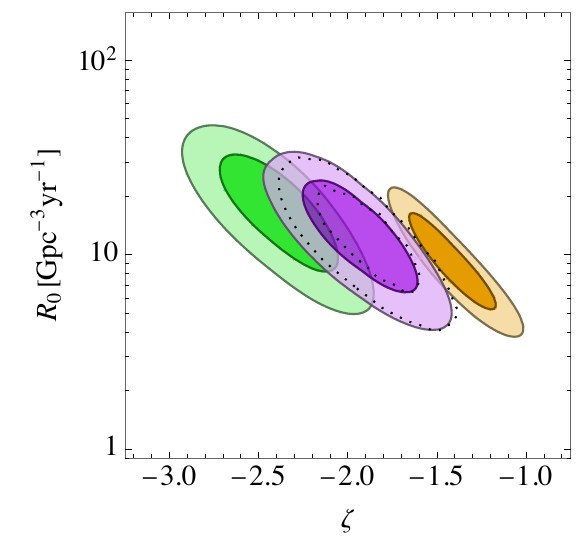} \hspace{2mm}
\includegraphics[height=0.4\textwidth]{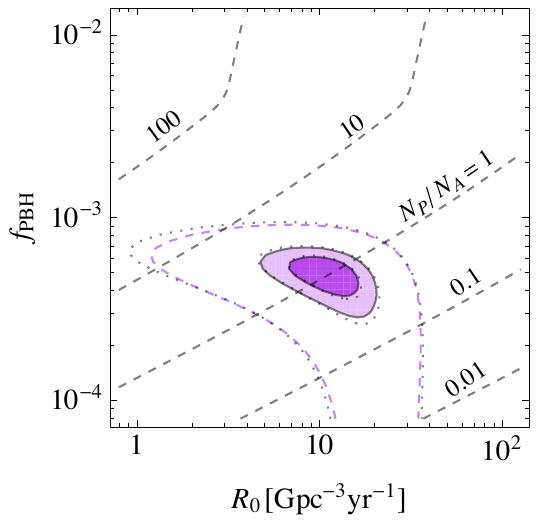}
\caption{The darker (lighter) shaded areas correspond to the $1\sigma$ ($2\sigma$) regions. {\emph Left panel:} The ABH merger rate, assuming a truncated power-law mass function, fitted to all LIGO-Virgo events (yellow) and to the selected events (green). The purple regions show the result of the combined fit of primordial and astrophysical BH merger rates assuming a log-normal PBH mass function. The dotted contours instead show the result of the combined fit assuming a critical collapse PBH mass function. {\emph Right panel:} PBH abundance in the maximum likelihood fit assuming log-normal PBH mass function and including astrophysical background modelled with a truncated power-law mass function. The dashed purple contour shows the boundary of the $5\sigma$ likelihood region. The grey dotted contours instead show the corresponding fit assuming a critical collapse PBH mass function.}
\label{fig:Afit}
\end{figure}

We assume a truncated power-law mass function for ABHs,~\cite{Abbott:2020gyp}
\be
    \psi(m) \propto m^{\zeta} \, \theta(m - m_{\rm min})\theta(m_{\rm max}- m) \,,
\ee
with the normalization $\int \psi(m) \td\ln m = 1$. We take $m_{\rm min} = 3.0 \Msun$ and $m_{\rm max} = 55 \Msun$ corresponding to an estimate for the maximal mass of neutron stars and the beginning of the PISN gap, respectively. We fix $\alpha=0$ and $\beta = 6$,\footnote{Without loss of generality we can take $\alpha = 0$ (\ie, we replace $\zeta \to \zeta - \alpha/2$ and $\beta \to \beta + \alpha/2$). By performing a three-parameter fit for $R_0$, $\zeta$ and $\beta$ we find that the fit is insensitive on the value of $\beta$.} and perform a maximum likelihood analysis for $R_0$, and $\zeta$. The result, assuming that all observed events are of astrophysical origin, is shown in Fig.~\ref{fig:Afit} by the yellow regions. The red regions instead show the fit to the low mass events depicted in the right panel of Fig.~\ref{fig:bestfitMF}. As the maximal mass $m_{\rm max}$ is not changed, this fit prefers steeper power-laws than the fit to all observed events.

Finally, we perform a combined fit where we sum the astrophysical and primordial BH merger rates and fit five parameters: $\langle m\rangle$, $\sigma_m$ and $f_{\rm PBH}$ parametrizing the PBH merger rate, and $\gamma$ and $R_0$ parametrizing the ABH merger rate. The $1\sigma$ and $2\sigma$ CL regions of the combined fit to all observed events are shown by in purple shaded regions in Figs.~\ref{fig:Pfit} and~\ref{fig:Afit}. The combined fit indicates an ABH merger rate in the range $6 - 20\,\Gpc^{-1}\yr^{-1}$. The rate is slightly larger for the ABH only fit and somewhat smaller than the fit to the light binary sub-population illustrated by the green histogram in the right panel of Fig.~\ref{fig:bestfitMF}.

The combined fit selects the PBH models for the heavier end of the mass distribution while low mass events are best fit by the ABH model as can be seen from Fig.~\ref{fig:bestfitMF}. Both the log-normal and the critical collapse mass functions fit the data almost equally well. This situation is very similar to the PBH only fits of Section~\ref{sec:implications} where the current data is mostly sensitive only to the mean $\langle m\rangle$ and the variation $\sigma_m$ of the PBH masses. In the combined fit, the expected number of merger events $N_{\rm P}$ from the PBH model is comparable to the number of merger events $N_{\rm A}$ from the ABH model, as is indicated by the black dashed contours in the right panel of Fig.~\ref{fig:Afit}.

Interestingly, the combined fit gives similar results as the fit using events divided into groups of light and heavy BHs by hand. The best-fit regions for the parameters $\langle m\rangle$ and $\sigma_m$ describing the shape of the PBH mass function overlap in Fig.~\ref{fig:Pfit}. At the same time, the PBH abundance is predicted to be slightly larger for the case where the events are separated by hand. This is because, unlike for the fit with separated events for which the ABH merger rate of BHs heavier than $15 \Msun$ is suppressed by construction, the ABH merger rate of heavier BH can be larger in the combined fit.

\begin{table}
\centering
\begin{tabular}{ l | l l l l l l l }
\hline\hline
      & A   & P, LN & P, CC  & C, LN    & C, CC   & S, LN     & S, CC  \\
\hline
$\ln(\max \ell_j/\max \ell_{\rm C, LN})$ & -6.2 & -13.6 & -13.8 & 0 & -1.6 & -5.3  & -7.4  \\
\hline\hline
\end{tabular}
\caption{Logarithm of the ratio of maximum likelihoods $\ln(\max \ell_j/\max \ell_{\rm C, LN})$ between different fits and the combined fit with a log-normal PBH mass function. The abbreviations are A=only ABHs, P=only PBHs, LN=log-normal PBH mass function, CC=critical collapse PBH mass function, S=separated events, C=combined fit.}
\label{table:likelihoods}
\end{table}

The likelihood ratios of the fits are summarized in Table~\ref{table:likelihoods}. We see that the combined fit gives a significant improvement compared to the fit including only ABHs. Moreover, the scenario where the present ABH merger rate is negligible, $R_0\approx0$ is strongly disfavoured. This is indicated by the small likelihood ratio $\ln[\max \ell_{\rm P}/\max \ell_{\rm CC}] \approx -13.6$, corresponding to about $4\sigma$ for a five parameter fit obeying Gaussian statistics. For the fit where the events are separated at $15\Msun$ as shown in the right panel of Fig.~\ref{fig:bestfitMF} the best-fit likelihood is still quite far from the one obtained in the combined fit. However, we stress that the separation of events is illustrative, and one should not directly compare the likelihood of the analysis with separated events to the one without such separation. Finally, the fit combining both astrophysical and primordial BH populations is slightly better than the astrophysical fit alone. Our finding that PBH only scenarios are disfavoured is in line with the earlier analysis of Ref.~\cite{Hall:2020daa} based on the LIGO-Virgo O1 and O2 observing runs.

Treating $m_{\rm max}$ as a free parameter, improves the log-likelihood of the fit including only ABHs so that $\ln(\max \ell_j/\max \ell_{\rm C, LN}) = -4.8$. The best-fit mass cut-off is $m_{\rm max} = 63\Msun$. However, this improvement is insignificant when penalizing for the increased number of parameters. This log-likelihood difference corresponds to the combined scenario being preferred by about $1.7\sigma$ for a five parameter fit obeying Gaussian statistics, which is in agreement with the LIGO-Virgo analysis that found a $1.4\sigma$ evidence for a peak in the mass distribution~\cite{Abbott:2020gyp}.

\section{Conclusions}
\label{sec:concl}

The origin of BH binaries responsible for the LIGO-Virgo GW events is currently unknown. In this paper, we analysed models of primordial and astrophysical BH binaries using the GWTC-2 dataset. We considered PBH binaries that are formed in the early Universe and whose mass distribution is characterised by log-normal and critical collapse mass functions. For ABHs, we assumed a simple truncated power-law mass distribution and performed a maximal likelihood analysis based on the mass and redshift dependence of the merger rates.

Our analysis revealed that the observed events are best fit by a model combining events from both astrophysical and primordial sources, while the PBH only scenarios were strongly disfavoured. However, due to the simplicity of the used ABH model, it is not possible to interpret our result as a hint for the PBH population in the LIGO-Virgo data. In fact, the preference for mass distributions with non-trivial features, such as peaks, was noted already in the LIGO-Virgo Collaboration paper~\cite{Abbott:2020gyp}, which is consistent with ours. Instead, our main finding is that the purely primordial scenario for the progenitor BHs is disfavoured, for both the log-normal and critical collapse mass functions. For both of these mass functions, the all fits select similar ranges for the mean mass and the width of the PBH mass distribution.

The mass functions preferred by the scenario in which all events are assumed to be primordial are centred around $19 \Msun$ with a mass dispersion of about $12 \Msun$ in both cases, while the required PBH abundance is $0.1\%$ for both mass functions considered. Including an astrophysical component increases the best-fit mean mass of the PBH distribution to $30 \Msun$ while the preferred mass dispersion remains roughly the same. The PBH abundance required to explain the signal dropped by 50\% indicating that the astrophysical component made up the other half of the merger rate. For the ABH+PBH scenario we find that at $1\sigma$ CL the ABH and PBH merger rates are given by $7 - 14\,\Gpc^{-1}\yr^{-1}$ and $0.4 - 1.4\,\Gpc^{-1}\yr^{-1}$, respectively. We find that vanishing ABH merger rate is disfavoured by $4\sigma$.

The LIGO-Virgo O3a results strengthen the constraints on the PBH abundance. The most conservative constraint arises when any combination of the observed GW events is allowed to be primordial, implying that $2-400\Msun$ PBHs must comprise less than 0.2\% of dark matter. For a monochromatic PBH mass function, the LIGO-Virgo observations exclude the scenario where all the DM is in PBHs in the mass range $0.4-800\Msun$. This mass range will, in general, be extended for broader mass functions. Additionally, the non-observation of the SGWB now excludes the scenario where more than $10\%$ of PBH is DM in the $2-100\Msun$ mass range, which is an almost one order of magnitude improvement over the previous constraint based on the LIGO-Virgo O2 run.


Due to large uncertainties in gas accretion physics in compact binary systems along with generic expectations that PBHs with moderate masses ${\cal O}(30)\Msun$ cannot grow their mass significantly, we neglected accretion effects and did not use the spin information in our analysis. Nevertheless, it is clear that spin statistics available from the upcoming larger and more sensitive datasets, in conjunction with detailed theoretical modeling, will undeniably become one of the most promising ways for discriminating various BH merger channels.

\acknowledgments 
We thank Marek Lewicki and Joosep Pata for useful discussions. This work was supported by the European Regional Development Fund through the CoE program grant TK133, the Mobilitas Pluss grants MOBTP135, MOBTT5, MOBTT86, and the Estonian Research Council grant PRG803. The work of VV is mainly supported by Juan de la Cierva fellowship from the Spanish State Research Agency. This work was also partly supported by the grants FPA2017-88915-P and SEV-2016-0588. IFAE is partially funded by the CERCA program of the Generalitat de Catalunya.

\appendix

\section{Approximate initial state suppression factor for the PBH merger rate}
\label{app:S1}

In this appendix we consider the approximation for the suppression factor $S_1$ given in Eq.~\eqref{eq:S1_appr}. The full suppression factor $S_1$ is~\cite{Raidal:2018bbj}
\be\label{eq:S1}
    S_1 = \frac{e^{-\bar{N}(y)}}{\Gamma(21/37)} \int \td v \, v^{-\frac{16}{37}} \exp\left[ -\bar{N}(y) \int \td \ln m \, \psi(m) F\left(\frac{m}{\langle m \rangle} \frac{v}{\bar{N}(y)}\right)  - \frac{3\sigma_{\rm M}^{2} v^{2}}{10 f_{\rm PBH}^{2}}  \right] \,,
\ee
where $F(z) = {}_1F{}_2\left(-1/2;3/4,5/4;-9z^2/16 \right) - 1$ with ${}_1F{}_2$ denoting the generalised hypergeometric function. In practical applications, the numerical evaluation of this double integral can be quite computationally expensive. An analytic approximation can be constructed by noting that the bounds on $S_1$,
\be
	S_{1,\rm min} \leq S_1 \leq S_{1,\rm max} < 1 \,,
\ee
can be obtained analytically
\bea
    S_{1,\rm max} 
&    = \left(\frac{5f_{\rm PBH}^{2}}{6 \sigma_{\rm M}^{2}}\right)^{\frac{21}{74}} U\left(\frac{21}{74},\frac{1}{2},\frac{5f_{\rm PBH}^{2}}{6 \sigma_{\rm M}^{2}}\right) 
\,,\\
	S_{1,\rm min}  
&	= \frac{\sqrt{\pi}(5/6)^{21/74}}{\Gamma(29/37)}  \left[\frac{\langle m^2\rangle/\langle m\rangle^2}{\bar{N}(y)} + \frac{\sigma_{\rm M}^2}{f_{\rm PBH}^2} \right]^{-\frac{21}{74}} e^{-\bar{N}(y)} \,,
\eea
where $U$ denotes the confluent hypergeometric function. These bounds are saturated in the limiting cases $\bar{N}(y) \to 0$ and $\bar{N}(y) \to \infty$, respectively. Thus we can approximate $S_1$ by extrapolating between $S_{1,\rm min}$ and $S_{1,\rm max}$ using a suitable ansatz for $S_1$ such as
\be\label{eq:S1_approx}
    S_1 \approx \frac{\sqrt{\pi}(5/6)^{21/74}}{\Gamma(29/37)}
    \left[\frac{\langle m^2\rangle/\langle m\rangle^2}{\bar{N}(y) + C } + \frac{\sigma_{\rm M}^2}{f_{\rm PBH}^2}\right]^{-\frac{21}{74}} e^{-\bar{N}(y)} \,,
\ee
and fixing $C$ by demanding that $S_1 \to S_{1,\rm max}$ in the limit $\bar{N}(y) \to 0$. This gives
\be
    C 
    =  f_{\rm PBH}^{2} \frac{\langle m^2\rangle/\langle m\rangle^2}{\sigma_{\rm M}^{2}} \left\{ \left[ \frac{\Gamma(29/37)} {\sqrt{\pi}} U\left(\frac{21}{74},\frac{1}{2},\frac{5f_{\rm PBH}^{2}}{6 \sigma_{\rm M}^{2}}\right) \right]^{-\frac{74}{21}} - 1 \right\}^{-1}.
\ee
We find that this approximation is accurate within 7\% for the log-normal mass function with $\sigma \leq 2$. 

Another approximation, tailored for the log-normal mass function, was constructed in Ref.~\cite{Hall:2020daa}. Although it is accurate within 2\%, it requires numerical integration over $v$ in Eq.~\eqref{eq:S1}. Given other theoretical uncertainties, \eg, uncertainties inherent to the non-linear evolution of the PBH surrounding the initial binary or uncertainties related to accretion physics, the approximation given in Eq.~\eqref{eq:S1_approx} is sufficient for our current purposes.

\begin{figure}
\centering
\includegraphics[width=0.98\textwidth]{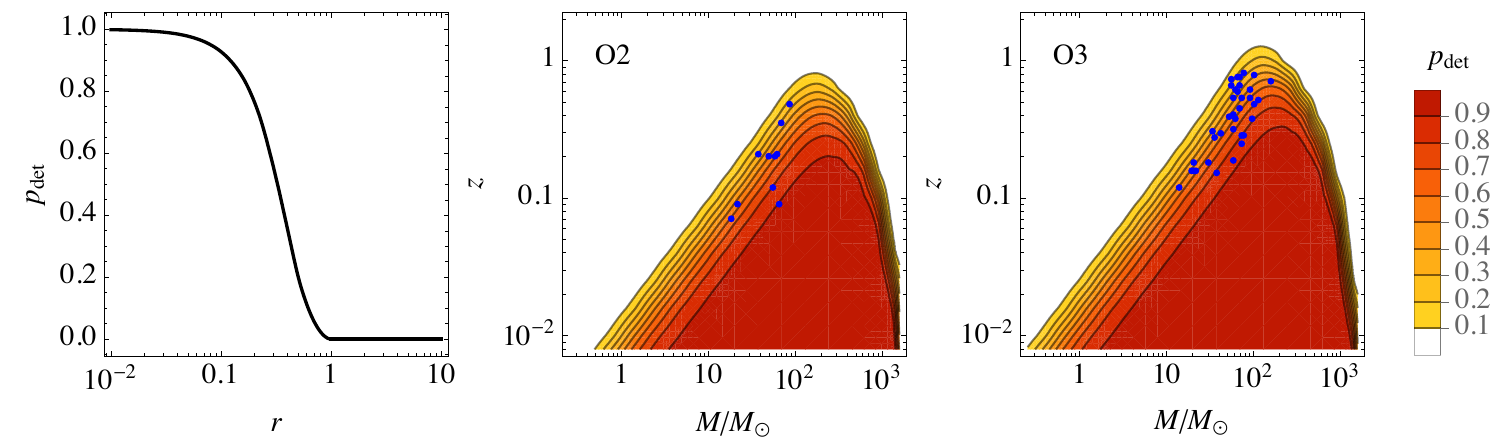}
\caption{The detection probability $p_{\rm det}$ as a function of $r={\rm SNR}_c/{\rm SNR}$ (left panel) and for equal mass binaries of total mass $M$ at redshift $z$ for O1\&O2 and O3a sensitivities (middle and right panels). The blue dots indicate the best-fit total masses and redshifts of the LIGO-Virgo BH mergers.}
\label{fig:pdet}
\end{figure}

\section{Detectability of BH binaries and the SGWB}
\label{app:detdectability}

We estimate the detection probability as~\cite{Finn:1992xs,Gerosa:2019dbe}
\be
p_{\rm det}(r) = \int_{r}^1 p(\omega) \td \omega
\ee 
with $r={\rm SNR}_c/{\rm SNR}$. The probability distribution $p(\omega)$ of the projection parameter $\omega\in[0,1]$ is defined as
\be
\omega^2 = \frac{(1+c_i^2)^2}{4} F_+(\theta,\phi,\psi)^2 + c_i^2 F_{\times}(\theta,\phi,\psi)^2 \,,
\ee
where the antenna functions $F_{+,\times}$ for a L-shaped detector are
\bea
&F_+(\theta,\phi,\psi) = \frac12 (1+\cos^2\theta) \cos2\phi \cos2\psi - \cos\theta \sin2\phi \sin2\psi \,, \\
&F_{\times}(\theta,\phi,\psi) = \frac12 (1+\cos^2\theta) \cos2\phi \sin2\psi + \cos\theta \sin2\phi \cos2\psi \,.
\eea
We compute $p(\omega)$ assuming uniform distributions for the binary inclination $c_i\in(-1,1)$, sky location $\cos\theta\in(-1,1)$ and $\phi\in(0,2\pi)$ and the polarization angle $\psi\in(0,2\pi)$. The left panel of Fig.~\ref{fig:pdet} shows the detection probability $p_{\rm det}$ as a function of $r$. The signal-to-noise ratio is defined as 
\be
	{\rm SNR} = \sqrt{4\int_0^\infty \td f\, \frac{|\tilde{h}(f)|^2}{S_n(f)}} \,,
\ee
where $\tilde{h}(f) = A(f) e^{i\Psi(f)}$ is the Fourier transform of the signal and $S_n(f)$ is the noise power spectrum of the GW detector, and its threshold value ${\rm SNR}_c$. Following Ref.~\cite{Ajith:2007kx}, we estimate the optimal amplitude $A(f)$ of the inspiral-merger-ringdown signal as\footnote{The amplitude is $A \propto \omega$, and in the optimal orientation $\omega=1$.} 
\be
A(f) = \sqrt{\frac{5\eta}{24}}\, \frac{\left[G M(1+z)\right]^{5/6}}{\pi^{2/3} D_L} \times
\begin{cases}
f^{-7/6}\,, & f<f_{\rm merg}\,, \\
f_{\rm merg}^{-1/2} f^{-2/3}\,, & f_{\rm merg} \leq f < f_{\rm ring} \,,\\
f_{\rm merg}^{-1/2} f_{\rm ring}^{-2/3} \frac{\sigma^2}{4(f-f_{\rm ring})^2 + \sigma^2}\,, & f_{\rm ring} \leq f < f_{\rm cut}\,, \\
\end{cases}
\ee
where $D_L$ is the luminosity distance of the source and the frequencies $f_{\rm merg}$, $f_{\rm ring}$, $f_{\rm cut}$ and $\sigma$ are of the form 
\be
    f_j = \frac{a_j \eta^2 + b_j \eta + c_j}{\pi GM(1+z)},
\ee 
with the coefficients $a_j$, $b_j$ and $c_j$ given in Table~I of Ref.~\cite{Ajith:2007kx}.

The BH binaries whose GW signal can not be resolved individually form a SGWB. The dimensionless energy density of this SGWB is\footnote{We include the factor $1-p_{\rm det}({\rm SNR}_c/{\rm SNR})$ that removes the events that can be seen individually. This affects the strength of the SGWB mostly at high PBH masses.}
\be \label{eq:GWB}
	\Omega_{\rm GW}(f) = \int \td \lambda \,\frac{1}{\rho_c} \frac{\td \rho_{\rm GW}}{\td f} \, \left[1-p_{\rm det}({\rm SNR}_c/{\rm SNR})\right]\,,
\ee
where $\rho_c$ denotes the critical density, and the GW energy density emitted by a binary in the frequency range $(f,f+\td f)$ is~\cite{Moore:2014lga}\footnote{Here the factor $4/25$ arises from averaging over the sky location of the binary and the orientation of its angular momentum (\ie, it is the expectation value of $\omega^2$).}
\be
    \td \rho_{\rm GW} = \frac{4}{25}\frac{\pi}{G} \,f^3 |\tilde{h}(f)|^2 \td f \,.
\ee
The detectability of the SGWB is determined by the signal-to-noise ratio 
\be
    {\rm SNR}_{\rm BG} = \sqrt{ \int_0^\mathcal{T} \td t \int \td f \left[\frac{\Omega_{\rm GW}(f)}{\Omega_n(f)}\right]^2} \,,
\ee
where $\Omega_n(f) = \pi f^3 S_n(f)/(4\rho_c)$ is the dimensionless energy density in noise~\cite{Moore:2014lga}.

For our analysis, we take the LIGO-Virgo O1\&O2 and O3a noise power spectra from Refs.~\cite{LIGOScientific:2018mvr,Abbott:2020niy} (see Fig.~\ref{fig:pdet} for the corresponding detection probabilities), and $\mathcal{T}_{\rm O2}=165\,$days and $\mathcal{T}_{\rm O3}=140\,$days for the observation times of O1\&O2 and O3a observations runs of LIGO-Virgo network. We then compute the network signal-to-noise ratio as the quadrature sum of the signal-to-noise ratios at individual interferometers, and use ${\rm SNR}_c = 8$ as the threshold signal-to-noise ratio.

\bibliography{PBH}
\end{document}